\documentclass[11pt,twoside,twocolumn]{article}

\usepackage{amsfonts,amsmath,amssymb,bm,abstract,balance,natbib,graphicx}

\usepackage[active]{srcltx}

\title{\textbf{Stable Langmuir solitons in plasma with diatomic ions}}

\author{Maxim Dvornikov$^{a,b}$
\\
$^{a}$\small{\ Institute of Physics, University of S\~{a}o Paulo,} \\
\small{CP 66318, CEP 05315-970 S\~{a}o Paulo, SP, Brazil} \\
$^{b}$\small{\ Pushkov Institute of Terrestrial Magnetism, Ionosphere} \\
\small{and Radiowave Propagation (IZMIRAN),} \\
\small{142190 Moscow, Troitsk, Russia;} \\
\small{E-mail: maxim.dvornikov@usp.br}}

\date{}

\begin{document}

\twocolumn[\maketitle
\begin{onecolabstract}
We study stable axially and spherically symmetric spatial solitons
in plasma with diatomic ions. The stability of a soliton against the
collapse is provided by the interaction of induced electric dipole
moments of ions with rapidly oscillating electric field of a plasmoid.
We derive the new cubic-quintic nonlinear Schr\"{o}dinger equation which
governs the soliton dynamics and numerically solve it. Then
we discuss the possibility of implementation of such plasmoids in realistic atmospheric plasma. In particular, we suggest that
spherically symmetric Langmuir solitons, described in the present work, can be excited at the formation stage
of long-lived atmospheric plasma structures. The implication of our model for the interpretation of the results of experiments for the plasmoids generation is discussed.
\end{onecolabstract}]

\maketitle

\section{Introduction\label{INTR}}

Stable spatial solitons are observed in the studies of optical phenomena~\citep{Seg98},
in solid states physics~\citep{Bur99}, and in the plasma research~\citep{Ant81}.
Typically a soliton can be stable owing to a certain nonlinearity.
For example, spatial plasma solitons, described in frames of the classical
electrodynamics, can exist due to the combined action of electron-ion
and electron-electron nonlinear interactions. The former one was found to be focusing~\citep{Zak72}, whereas the latter interaction
can be defocusing~\citep{Kuz76,SkoHaa80,DavYakZal05}. Recently \citet{HaaShu09} suggested
that, if one accounts for the additional quantum pressure of electron gas,
it may explain the appearance of two and three dimensional Langmuir solitons in dense plasmas.

In the present work we stall study the existence of stable Langmuir
solitons in a plasma with ions possessing induced electric dipole
moments (EDM). We shall demonstrate that the interaction of ion's
EDM with a rapidly varying electric field of a plasma oscillation
results in the arrest of the Langmuir wave collapse. Thus the existence
of a spatial soliton becomes possible.

In our analysis we shall choose a plasma with diatomic nitrogen ions, which
corresponds to a realistic atmospheric plasma. Thus our results
may be applied for the theoretical description of long-lived plasma
structures observed in the atmosphere~\citep{BycNikDij10}. It is interesting
to notice that the role of EDM of charged particles for the explanation
of the stability of atmospheric plasmoids was also discussed previously by~\citet{Ber73,Sta73}.

This work is organized as follows. In Sec.~\ref{sec:NLLO} we consider
the general description of nonlinear waves in plasma. We introduce
the new ponderomotive force, associated with EDM, which acts on
the ion component of plasma. Then we derive a system of nonlinear
equations for the electric field amplitude and the perturbation of
the ion density. In Sec.~\ref{sec:CQNLSE} we reduce this system
to a single nonlinear Schr\"{o}dinger equation (NLSE),
containing cubic and quintic terms, for the envelope
of the electric field. This equation
is analyzed numerically for the case of a radial plasma oscillation.
We show that both axially and spherically symmetric stable solitons
can exist. Then, in Sec.~\ref{sec:APPL}, we discuss a possible
application of our model of stable spatial solitons to the description
of long-lived atmospheric plasmoids as well as for the interpretation of some experimental results. Finally, in Sec.~\ref{sec:CONCL},
we briefly summarize our results. The calculation of the permittivity
of the ion component of plasma is presented in Appendix~\ref{sec:POLARIZ}.

\section{\label{sec:NLLO}Nonlinear Langmuir oscillations in plasma}

In this section we shall derive the basic nonlinear equations for
the description of Langmuir waves in plasma accounting for a ponderomotive
force acting on nonpolar diatomic ions.

If we study electrostatic plasma oscillations, i.e. when the magnetic
field is zero, $\mathbf{B} = 0$, the motion of the electron component of plasma obeys the
following plasma hydrodynamics equations:
\begin{align}
  \frac{\partial n_{e}}{\partial t} + \nabla\cdot(n_{e}\mathbf{v}_{e}) = & 0,\nonumber \\
  \frac{\partial\mathbf{v}_{e}}{\partial t} + (\mathbf{v}_{e}\nabla)\mathbf{v}_{e} = &
  -\frac{e}{m}\mathbf{E} - \frac{1}{n_{e}m}\nabla p,\label{eq:electrhyd}
\end{align}
where $n_{e}$ is the number density of electrons, $\mathbf{v}_{e}$
is the electron velocity, $m$ is the mass of an electron, $e>0$
is the proton charge, $\mathbf{E}$ is the strength of the
electric field, and $p$ is the pressure. We should also consider Maxwell and Poisson equations for the electric field evolution,
\begin{align}\label{MaxPoi}
  \frac{\partial\mathbf{E}}{\partial t} = & 4\pi e (n_{e}\mathbf{v}_{e} - n_{i}\mathbf{v}_{i}),\nonumber \\
  (\nabla\cdot\mathbf{E}) = & -4\pi e (n_{e}-n_{i}),
\end{align}
where $n_{i}$ is the ion number density and $\mathbf{v}_{i}$ is
the ion velocity.

In the zeroth approximation only electrons participate in a plasma
oscillation, with the number density of ions being approximately constant:
$n_{i}\approx n_{0}=\text{const}$. Thus we may present the electric
field in the form,
\begin{equation}\label{eq:ELangosc}
  \mathbf{E} = \mathbf{E}_{1} e^{-\mathrm{i}\omega_{e}t} +
  \mathbf{E}_{1}^{*}e^{\mathrm{i}\omega_{e}t}+\dotsb,
\end{equation}
where $\omega_{e}=\sqrt{4\pi e^{2}n_{0}/m}$ is the Langmuir frequency
for electrons and $\mathbf{E}_{1}$ is the amplitude of the electric field. It should be noted that in the following (see, e.g., Sec.~\ref{sec:CQNLSE}) we shall study axially and spherically symmetric plasma oscillations. In this case one can find a scalar potential $\phi_1$ such as $\mathbf{E}_{1} = - \nabla \phi_1$ in Eq.~\eqref{eq:ELangosc}.

In a realistic situation ions will also participate in a plasma oscillation.
Thus the ion density becomes $n_{i}=n_{0}+n(\mathbf{r},t)$, where
$n$ is the perturbation of the ion density. It leads to the appearance of higher harmonics omitted
in Eq.~\eqref{eq:ELangosc}. The plasma hydrodynamic equations for the description
of the ions evolution have the form~\citep{SkoHaa80},
\begin{align}
  \frac{\partial n_{i}}{\partial t} + \nabla\cdot(n_{i}\mathbf{v}_{i}) = & 0,\nonumber \\
  \frac{\partial\mathbf{v}_{i}}{\partial t} + (\mathbf{v}_{i}\nabla)\mathbf{v}_{i} = &
  \frac{\mathbf{F}_i}{M}, %\nabla\varphi_{\mathrm{eff}},%+\frac{1}{n_{i}M}\mathbf{f}_{\mathrm{pol}},
  \label{eq:ionhyd}
\end{align}
where $M$ is the ion mass, $\mathbf{F}_i$ is the force acting of ions. The reason why one can omit the ion pressure term in Eq.~\eqref{eq:ionhyd} will be discussed at the end of this section.

Using the quasineutrality of plasma we can find that~\citep{Zak72}
\begin{equation}\label{quasineu}
    n_0 + n \approx n_e = n_0
    \exp
    \left(
      \frac{e\varphi_s - U_{\mathrm{pm}}}{T_e}
    \right)
\end{equation}
where $\varphi_s$ is the slowly varying part of the electric potential,
$T_{e}$ is the electron temperature, and $U_{\mathrm{pm}}=|\mathbf{E}_{1}|^{2}/(4\pi n_0)$
is the potential of the ponderomotive force which acts on a charged
particle in a rapidly oscillating electric field given in Eq.~\eqref{eq:ELangosc}. Supposing that ions are mainly involved in the slow motion of plasma we get that $\mathbf{F}_i = - e \nabla \varphi_s$. Finally, using Eqs.~\eqref{eq:electrhyd}-\eqref{quasineu} one arrives to the system of equations of~\citet{Zak72}. More detailed derivation of the nonlinear plasma evolution equations which include electron-ion and electron-electron interactions can be found in the works by~\citet{Kuz76,SkoHaa80}.

It should be noted that Eq.~\eqref{eq:ionhyd} is derived
under the assumption of ions having point like charges. However
realistic atmospheric plasma contains mainly nitrogen or oxygen ions, which
are diatomic (see also Sec.~\ref{sec:APPL}). In this case the simplified
ion hydrodynamics Eq.~\eqref{eq:ionhyd}
%, which accounts for only the ponderomotive potential $U_{\mathrm{pm}}$,
is incomplete since it does not take
into account the internal structure of ions.

A diatomic molecule is nonpolar, i.e. it cannot have an intrinsic EDM because of the symmetry reasons.
Nevertheless, this kind of molecules can acquire EDM, $p_{i}=\alpha_{ij}E_{j}$, in an external
electric field. Here $(\alpha_{ij})$ is the polarizability tensor. Hence the additional force,
$\mathbf{F}_\mathrm{pol} = (\mathbf{p} \nabla) \mathbf{E}$, will act on this particle placed
in an external inhomogeneous electric field. Thus, if we study the plasma with diatomic ions, in Eq.~\eqref{eq:ionhyd} one should
replace $\mathbf{F}_i = - e \nabla \varphi_s \to - e \nabla \varphi_s + \mathbf{f}_\mathrm{pol}/n_i$,
where $\mathbf{f}_{\mathrm{pol}}$ is the volume density of the ponderomotive force
related to the matter polarization.

If an ion is diatomic and possesses an axial symmetry,
one can always reduce the polarizability tensor to the diagonal form, $(\alpha_{ij})=\mathrm{diag}\left(\alpha_{\bot},\alpha_{\bot},\alpha_{\Vert}\right)$,
where $\alpha_{\bot}$ and $\alpha_{\Vert}$ are transversal and longitudinal
polarizabilities. Now the expression for $\mathbf{f}_{\mathrm{pol}}$ can be obtained using Eq.~\eqref{eq:perm} and the general technique developed by~\citet{Tam79}, as
\begin{align}\label{eq:fpmpol}
  \mathbf{f}_{\mathrm{pol}} = & \frac{1}{8\pi}
  \left[
    \nabla
    \left(
      n_{i}
      \frac{\partial\varepsilon}{\partial n_{i}}
      \mathbf{E}^{2}
    \right) -
    \mathbf{E}^{2}\nabla\varepsilon
  \right]
  \notag
  \\
  & =
  n_{i}
  \left[
    \langle\alpha\rangle+\frac{4}{45}
    \frac{(\Delta\alpha\mathbf{E})^{2}}{T_{i}}
  \right]
  \nabla\mathbf{E^{2}},
\end{align}
where $\varepsilon$ is the permittivity of the ion component of plasma,
$T_{i}$ is the ion temperature, $\langle\alpha\rangle=(2\alpha_{\perp}+\alpha_{\parallel})/3$
is the mean polarizability of an ion, and $\Delta\alpha=\alpha_{\parallel}-\alpha_{\perp}$.

It should be noted that the general expression for the ponderomotive force $\mathbf{f}_{\mathrm{pol}}$
was derived by~\citet{Tam79} under the assumption of static fields with
$(\nabla \times \mathbf{E}) = 0$.
However, as we mentioned above, we study electrostatic plasma oscillations with zero magnetic field
(see also Sec.~\ref{sec:CQNLSE}). Thus Eq.~\eqref{eq:fpmpol} remains valid.

Combining Eqs.~\eqref{eq:electrhyd}-\eqref{eq:fpmpol} we get the following
nonlinear coupled equations for the amplidute of the electric field,
\begin{equation}\label{eq:ZakheqE}
  \mathrm{i}\dot{\mathbf{E}}_1 +
  \frac{3}{2}\omega_{e}r_{\mathrm{D}}^2\nabla(\nabla \cdot \mathbf{E}_1) -
  \frac{\omega_{e}}{2n_{0}}n\mathbf{E}_1 = 0,
\end{equation}
and for the perturbation of the ion density,
\begin{align}\label{eq:Zakheqn}
  \left(
    \frac{\partial^{2}}{\partial t^{2}}-c_{s}^{2}\nabla^{2}
  \right)
  n = &
  \frac{1}{4\pi M} \nabla^{2} |\mathbf{E}_1|^{2}
  \notag
  \\
  & -
  \frac{4}{15}
  \frac{(\Delta\alpha)^{2}n_{0}}{MT_{i}}\nabla^{2}|\mathbf{E}_1|^{4},
\end{align}
where $r_{\mathrm{D}}=\sqrt{T_{e}/4\pi e^{2}n_{0}}$ is the Debye length and $c_{s}=\sqrt{T_{e}/M}$ is the sound velocity in plasma. To derive
Eq.~\eqref{eq:Zakheqn} we take into account that $\langle\mathbf{E}^{4}\rangle=6|\mathbf{E}_{1}|^{4}$
while averaging over the time interval $\sim1/\omega_{e}$.

The first quadratic term in the rhs of Eq.~\eqref{eq:Zakheqn} corresponds to the direct interaction of charged ions
with the electric field whereas the second quartic term there, $\sim\nabla^{2}|\mathbf{E}_1|^{4}$,
is related to the induced EDM interaction. Hence the contribution of
this second term to the Langmuir waves dynamics should be typically smaller.
However, as we shall see in Sec.~\ref{sec:CQNLSE}, in some cases it is the EDM term which arrests the
collapse of Langmuir waves.

It should be noted that in Eq.~\eqref{eq:Zakheqn} we neglect the
contribution of the ion temperature to the sound velocity. Such a contribution would correspond to a nonzero ion pressure term in Eq.~\eqref{eq:ionhyd}. Since we suppose that
$T_{i}\ll T_{e}$, we can omit the ion pressure. However we keep the ion temperature in the quartic nonlinear
term in Eq.~\eqref{eq:Zakheqn}. In the rhs of Eq.~\eqref{eq:Zakheqn}
we also neglect term
$\sim - n_{0}\langle\alpha\rangle\nabla^{2}|\mathbf{E}_1|^{2}/M$,
which is small compared to the contribution of the Miller force. Indeed, the ratio of these terms is $\sim n_{0}\langle \alpha \rangle$. In Sec.~\ref{sec:APPL} we shall use the following values of $n_0$ and $\langle \alpha \rangle$: $n_0 \sim 10^{21}\thinspace\text{cm}^{-3}$ and $\langle \alpha \rangle \sim 10^{-24}\thinspace\text{cm}^{-3}$. For such a parameters, this ratio $\sim 10^{-3}$, that justifies the validity of Eq.~\eqref{eq:Zakheqn}.

Finally we notice that the quartic nonlinear term in Eq.~\eqref{eq:Zakheqn} blows up if $T_i \to 0$. It means that our model is not valid at the very low ion temperature. However, in Sec.~\ref{sec:APPL}, where we shall discuss a possible application, we consider only reasonable, relatively high values of $T_i$. Note that at extremely low temperatures quantum corrections to the ion and electron motion become important.

\section{\label{sec:CQNLSE}Cubic-quintic nonlinear Schr\"{o}dinger equation}

In this section we shall derive the nonlinear Schr\"{o}dinger equation
for the amplitude of the electric field. This equation will be analyzed in
a particular case of a radial plasma oscillation. We shall numerically
solve it and find the characteristics of Langmuir solitons.
The soliton stability will be examined.

Let us suggest that the density variation in Eq.~\eqref{eq:Zakheqn}
is slow, i.e. $\partial^{2}n/\partial t^{2}\ll c_{s}^{2}\nabla^{2}n$.
In this subsonic regime Eqs.~\eqref{eq:ZakheqE} and~\eqref{eq:Zakheqn}
can be cast in a single NLSE,
\begin{align}\label{eq:NLSEE}
  \mathrm{i}\dot{\mathbf{E}} & + \frac{3}{2}\omega_{e}r_{\mathrm{D}}^{2}\nabla(\nabla\cdot\mathbf{E})
  \notag
  \\
  & +
  \frac{\omega_{e}}{T_{e}}
  \left(
    \frac{1}{8\pi n_{0}}|\mathbf{E}|^{2}-
    \frac{2(\Delta\alpha)^{2}}{15T_{i}}|\mathbf{E}|^{4}
  \right)
  \mathbf{E}=0,
\end{align}
which has both cubic and quintic nonlinear terms. NLSEs analogous
to Eq.~\eqref{eq:NLSEE} were previously examined by~\citet{DesMaiMal00} in connection to
the studies of the light bullet propagation in crystals.
Note
that in Eq.~\eqref{eq:NLSEE} we omit the
index ``1'' in the amplitude of the electric field, i.e. $\mathbf{E}_{1}\equiv\mathbf{E}$,
in order not to encumber the formulas.

As we mentioned in Sec.~\ref{sec:NLLO}, we shall examine axially or spherically symmetric plasma oscillations,
i.e. $\mathbf{E}=E\mathbf{e}_{r}$, where $\mathbf{e}_{r}$ is a unit
vector in radial direction and $E$ is a scalar function. Introducing
the following dimensionless variables:
\begin{align}
  \tau = & \frac{15}{128\pi^{2}}
  \frac{T_{i}}{T_{e}}
  \frac{1}{(n_{0}\Delta\alpha)^{2}}
  \omega_{e}t,
  \notag
  \\
  x = & \frac{1}{8\pi n_{0}\Delta\alpha}
  \sqrt{\frac{5T_{i}}{T_{e}}}
  \frac{r}{r_{\mathrm{D}}},
  \notag
  \\
  \psi = & 4\Delta\alpha
  \sqrt{\frac{\pi n_{0}}{15T_{i}}}E,
  \label{eq:dimlessvar}
\end{align}
we can represent Eq.~\eqref{eq:NLSEE} in the form,
\begin{align}\label{eq:NLSEpsi}
  \mathrm{i}\frac{\partial\psi}{\partial\tau} & +
  \psi'' + \frac{d-1}{x}\psi' - \frac{d-1}{x^{2}}\psi 
  \notag
  \\
  & +
  \left(
    |\psi|^{2}-|\psi|^{4}
  \right)
  \psi = 0,
\end{align}
which contains no dimensionless parameters. Here $d=2,3$ is the dimension
of space.

One can check by the direct calculation that the plasmon number
\begin{equation}%\label{eq:plnum}
  N = \int_{0}^{\infty}\Omega_{d}\mathrm{\ d}x\ x^{d-1}|\psi|^{2},
\end{equation}
and the Hamiltonian
\begin{align}%\label{eq:Ham}
  H = & \int_{0}^{\infty}\Omega_{d}\mathrm{\ d}x\ x^{d-1}
  \notag
  \\
  & \times
  \bigg\{
    \left|
      \frac{1}{x^{d-1}}
      \left(
          x^{d-1}\psi
      \right)'
    \right|^{2} 
    \notag
    \\
    & -
    \frac{1}{2}|\psi|^{4}+\frac{1}{3}|\psi|^{6}
  \bigg\},
\end{align}
are the integrals of Eq.~\eqref{eq:NLSEpsi}. Here
$\Omega_{2}=2\pi$ and $\Omega_{3}=4\pi$ are the solid angles in
two and three dimensions.

We shall look for the solution of Eq.~\eqref{eq:NLSEpsi} as $\psi(x,\tau)=e^{\mathrm{i}\lambda\tau}\psi_{0}(x)$,
where $\lambda$ is a real number meaning the dimensionless frequency
shift. By the proper choice of the phase we can always make the function
$\psi_{0}$ to be real. The corresponding ordinary differential equation
for the function $\psi_{0}$ is solved numerically using the MATLAB
program.
%It requires an appropriate initial guess function $\psi_{g}(x)$.
%The guess function is taken to have the Gaussian form $\psi_{g}=Ax\exp(-x^{2}/2\sigma^{2})$,
%with the parameters $A$, $\sigma$, and $\lambda$ chosen to
%minimize $H$ at the constant $N$.

\begin{figure}
  \centering
  \includegraphics[scale=.9]{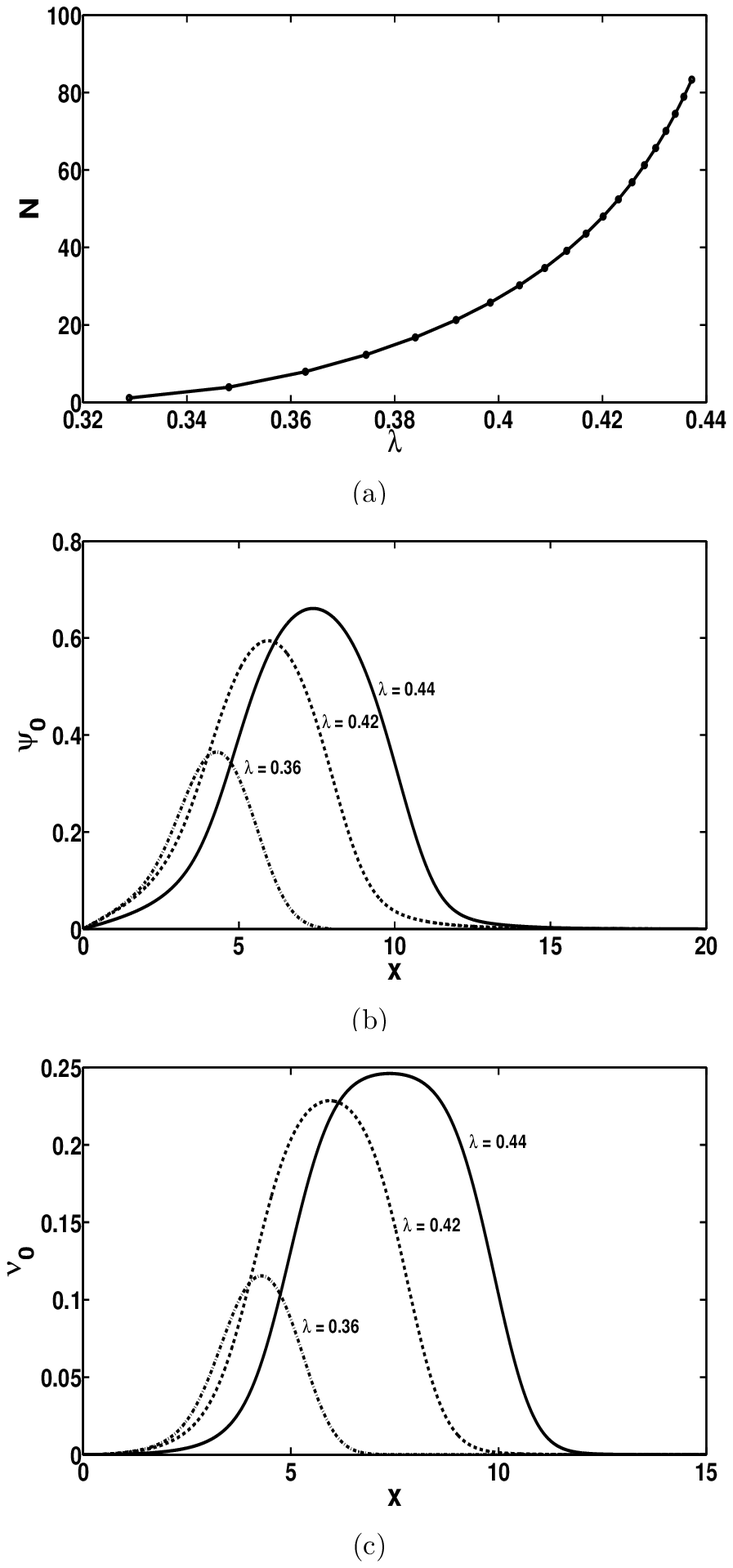}
  \caption{\label{fig:2Dsol}
  The analysis of Eq.~\eqref{eq:NLSEpsi} in 2D case. (a) The function
  $N(\lambda)$. (b) Examples of the numerical solitons $\psi_{0}(x)$ for different
  $\lambda$. (c) The function $\nu_0(x) = |\psi_0|^2 - |\psi_0|^4$ for different values of $\lambda$.}
\end{figure}

\begin{figure}
  \centering
  \includegraphics[scale=.9]{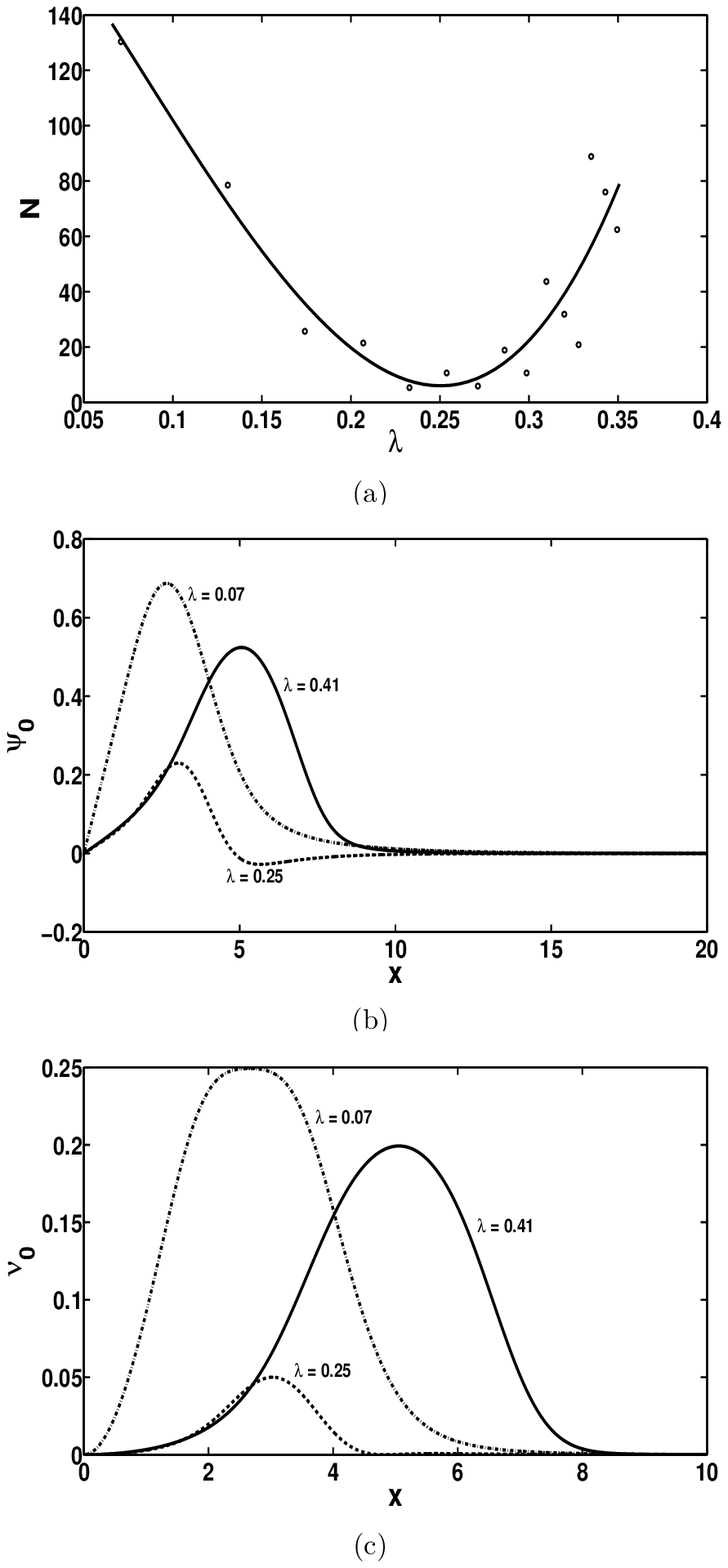}
  \caption{\label{fig:3Dsol}
  Panels (a)-(c) are the same as in Fig.~\ref{fig:2Dsol} but correspond
  to 3D case.}
\end{figure}

Firstly, we analyze the stability of axially and spherically symmetric
solitons by plotting $N(\lambda)$ dependence. It is shown in Fig.~\ref{fig:2Dsol}(a)
for 2D case and in Fig.~\ref{fig:3Dsol}(b) in 3D case. One should
notice that in 2D case $\partial N/\partial\lambda>0$ in a quite
broad range of $\lambda$. Thus according to \citet{KuzRubZak86} criterion
(VKC), this kind of 2D solitons is stable.
In 3D case the accuracy of calculations is significantly lower than
that in 2D situation. To build a smooth $N(\lambda)$ curve in Fig.~\ref{fig:3Dsol}(a)
the least squares method was used since the points on this plot, obtained
with numerical simulations in MATLAB, have rather big spread, especially
at large $\lambda$. Meanwhile one can see that unstable and stable
solitons coexist in 3D case. In Fig.~\ref{fig:3Dsol}(a) we get that
$\partial N/\partial\lambda<0$ at $\lambda\lesssim0.26$. Applying
VKC we conclude that this branch corresponds to unstable solitons.
However, if $\lambda\gtrsim0.26$, we obtain that $\partial N/\partial\lambda>0$,
that corresponds to stable solitons.

\section{\label{sec:APPL}Application}

In this section we discuss the application of the results obtained
in Secs.~\ref{sec:NLLO} and~\ref{sec:CQNLSE} for the studies of a natural atmospheric plasmoid called a \emph{ball
lightning} (BL).

BL is a glowing object appearing mainly during a thunderstorm~\citep{Keu13}. Despite
the existence of numerous BL models~\citep{BycNikDij10},
it is likely to be a plasma based phenomenon.
%The most spectacular BL
%property is its long life-time, up to several minutes, which is untypical
%for an unstructured plasma.
\citet{DvoDvo06,Dvo11,Dvo12PRSA,Dvo12JASTP,Dvo13} put
forward a hypothesis that BL is based on a radial oscillation of plasma and
discussed both classical and quantum approaches to this problem. Note that analogous BL model was also considered by~\citet{Fed99,Shm03,Ten11}.

Note that, in the BL model based on radial plasma oscillations, it is not required to have any special force which maintains the shape of a plasma structure. On the contrary, in case of a model based on the static distribution of electric charges inside a plasmoid one needs a self-consistent interaction, e.g., of the magnetic origin, which sustains a plasma object in equilibrium. However, in this case plasma will radiate electromagnetic waves and a plasmoid will lose its energy within the millisecond time scale.

We already mentioned in Sec.~\ref{sec:NLLO} that the magnetic field in the system of radially oscillating particles is equal to zero. This fact may explain the relative stability of a plasma structure since it does not lose energy by the electromagnetic radiation. Moreover, here we are using the concept of collisionless plasma. Later in this section we discuss how one can account for the visible radiation of BL described in frames of our model.

%The observed glow of BL may be accounted for by some peripheral effects, e.g., by small deviations of a plasma oscillation from its spherical shape. The glowing of BL puts certain limits on the life time of a plasmoid since it is related to the weak energy leakage from the system.

%It is interesting to mention that spherically and axially (snake-like
%BL, see Ref.~\cite{BycNikDij10p204}) symmetric oscillations of charged particles
%do not have a magnetic field. Thus this system do not lose energy
%by electromagnetic radiation.

%If we assume that BL is a spherical nonlinear plasma oscillation, we should suggest
%a mechanism to arrest the collapse of a Langmuir wave~\cite{Zak72}, to explain
%the BL stability. In the present work we described the mechanism for preventing
%the collapse based on the interaction of the ions EDM with the oscillating electric field of a plasmoid.

Let us discuss the conditions for the existence of a spherical nonlinear Langmuir soliton in the atmospheric plasma.
Firstly, an atmospheric plasma can contain $\mathrm{N}_{2}^{+}$ ($\approx78\%$) and $\mathrm{O}_{2}^{+}$
($\approx21\%$) ions. These ions are diatomic and nonpolar. We have already mentioned in
Sec.~\ref{sec:NLLO} that diatomic ions cannot have an intrinsic EDM, whereas an induced EDM
is well possible. Thus our results may be applied for the description of atmospheric plasmoids.

Now let us evaluate the characteristics of the plasma structure.
For the stability of a soliton its size should be greater than $r_{\mathrm{D}}$
-- otherwise thermal electrons will cross the soliton leading to the
development of a turbulence. One can see in Figs.~\ref{fig:2Dsol}(b) and~\ref{fig:3Dsol}(b)
that in dimensionless variables the typical soliton size can be up
to $10$. Using the values of the polarizabilities
of a \emph{neutral} nitrogen molecule~\citep{AlmBurFly75}, $\alpha_{ij}\sim10^{-24}\ \text{cm}^{-3}$,
and Eq.~\eqref{eq:dimlessvar}, we get that if
\begin{equation}\label{eq:xidef}
  \kappa =
  \left(
    \frac{n_0}{10^{21}\ \text{cm}^{-3}}
  \right)^{2}
  \frac{T_e}{T_i} > 79,
\end{equation}
a plasmoid becomes stable.

%for $n_{0} \gtrsim 10^{20}\ \text{cm}^{-3}$,
%For a rough estimate we shall take the
%typical ion temperature $T_{i}=300\thinspace\text{K}$ and suppose
%that the electron temperature in plasma is $T_{e}=10^{6}\ \text{K}$.

%This value is close to the density
%of an ideal gas at standard temperature and pressure, the Loschmidt's
%number, $n_{l}=2.7\times10^{19}\ \text{cm}^{-3}$.

The analysis of the soliton stability is based on the perturbation
theory which requires that $\delta\omega < \omega_{e}$, where $\delta\omega$ is the frequency shift.
Using the parameter $\kappa$ defined in Eq.~\eqref{eq:xidef} we get that this condition is equivalent to,
\begin{equation}\label{eq:perttval}
  \kappa > 4.7 \times 10^3,
\end{equation}
where we use Figs.~\ref{fig:2Dsol}(a) and~\ref{fig:3Dsol}(a) to obtain the typical value of $\lambda \sim 0.4$.

%\begin{equation}\label{eq:perttval}
%  \frac{\delta\omega}{\omega_{e}} =
%  \lambda\frac{15}{128\pi^{2}}
%  \frac{T_{i}}{T_{e}}
%  \frac{1}{(n_{0}\Delta\alpha)^{2}} < 1,
%\end{equation}
%
%Using the above chosen parameters of plasma and
%we get that the condition~\eqref{eq:perttval} is satisfied if $n_{0}\gtrsim10^{21}\ \text{cm}^{-3}$.
%which is one order of magnitude bigger than the previous constraint
%on $n_{0}$. Nevertheless, one should notice that the nonlinear ponderomotive
%force $\mathbf{f}_{\mathrm{pol}}$ in Eq.~\eqref{eq:ionhyd} will be
%defocusing in any case, providing the soliton stability. If condition~\eqref{eq:perttval}
%is violated, one should just analyze the complete set of nonlinear
%hydrodynamic equations for plasma rather than use the simplified perturbative
%treatment.

The ion density inside a plasmoid is lower than the ambient density~\citep{Zak72}. Integrating Eq.~\eqref{eq:Zakheqn} in the subsonic regime and using Eq.~\eqref{eq:dimlessvar} we should impose the constraint on the normalized ion density variation,
\begin{equation}\label{eq:nniondens}
  \left|
    \frac{n}{n_0}
  \right| \approx
  2.4 \times 10^4 \frac{\nu_0}{\kappa} < 1,
\end{equation}
where $\nu_0(x) = |\psi_0|^2 - |\psi_0|^4$. Note that $n$ is Eq.~\eqref{eq:nniondens} is negative. The function $\nu_0(x)$, corresponding to different numerical solutions of Eq.~\eqref{eq:NLSEpsi}, is shown in Figs.~\ref{fig:2Dsol}(c) and~\ref{fig:3Dsol}(c) for axial and spherical plasma structures. As one can see in these plots, the maximal value of $\nu_0$ is in the $(0.05 - 0.25)$ range.

In 2D case, the bigger the parameter $\lambda$ is, the greater the maximum of $\nu_0$ is. Thus, if we take $\lambda \lesssim 0.4$ and $\kappa > 6 \times 10^3$, then using Eq.~\eqref{eq:nniondens}, we get that $|n|<n_0$ inside the whole plasmoid. In 3D case, one can see in Fig.~\ref{fig:3Dsol}(c) that the smallest value of the maximum of $\nu_0(x)$ is archived for $\lambda_\mathrm{min} \approx 0.25$. Note that $\lambda_\mathrm{min}$ corresponds to the minimum of the function $N(\lambda)$, cf. Fig.~\ref{fig:3Dsol}(a). Again considering that $\lambda$ is close to $\lambda_\mathrm{min}$, one obtains that the ion number density is positive in a spherical plasma structure if $\kappa > 6 \times 10^3$.

The conditions of the plasma structure stability in Eqs.~\eqref{eq:xidef}-\eqref{eq:nniondens} are satisfied if we suppose that $n_{0} \sim 10^{21} \thinspace \text{cm}^{-3}$, $T_{i} = 300\thinspace\text{K}$ and $T_{e}=10^{6}\thinspace\text{K}$. Now let us estimate the typical size of a plasmoid requiring that
$R_{\mathrm{eff}}>r_{\mathrm{D}}$. For the chosen electron density and temperatures we get the lower bound for the radius: $R_{\mathrm{eff}} \gtrsim (10^{-7} \div 10^{-6})\ \text{cm}$. Using the numerical solutions of Eq.~\eqref{eq:NLSEpsi}, shown on Figs.~\ref{fig:2Dsol}(b) and~\ref{fig:3Dsol}(b), as well as obtained estimates for $\lambda$, we get the upper bound for the radius: $R_{\mathrm{eff}} \lesssim 10^{-5}\ \text{cm}$. Note that at $n_{0} \sim 10^{21}\ \text{cm}^{-3}$ and $R_{\mathrm{eff}} \sim 10^{-5}\ \text{cm}$, still there is a significant number of charged particles inside the plasmoid: $\tfrac{4}{3} \pi R_{\mathrm{eff}}^3 n_0 \sim 10^6$. It should be also noticed that the plasmoid radius is at least one order of magnitude greater than the intermolecular distance: $R_{\mathrm{eff}} \gg n_0^{-1/3}$.

It should be mentioned that the obtained small size of a plasma structure is very close to the estimates obtained by~\citet{DvoDvo06,Dvo12PRSA,Dvo12JASTP,Dvo13}, where radial nonlinear plasma oscillations were studied using the quantum approach. Note that a theoretical model of BL having a nano-sized kernel was developed by~\citet{Ale08} using the methods of generalized quantum hydrodynamics. The existence of plasma structures having the size, which lies in the nanometer range, was theoretically suggested by~\citet{AbrDin00}. Recently their hypothesis was experimentally tested by~\citet{DikJer06,Pai07}.
%The results of these experiments were analyzed by~\citet{SteMas08}.

According to observations, the visible dimension of BL is about several centimeters~\citep{Doe13}. However its core should be much smaller. Otherwise it is difficult to explain how BL passes through small holes or cracks in dielectric materials without destroying them~\citep{Doe13}. Therefore the visible size of a plasmoid is likely to be caused by some auxiliary effects.

We can also assume that BL is a composite object. This assumption is based on the fact that sometimes BL can decay into pieces~\citep{BycNikDij10}. The composite structure of plasma objects was also confirmed experimentally. The photographs taken by~\citet{Pai07} show that the visible size of artificial plasmoids lies in the cm range. However, the analysis of materials which were in contact with these plasma objects demonstrates that plasmoids consist of multiple small kernels. The models of a composite BL were previously discussed by~\citet{Mes07,Dvo12JASTP}. It should be noted that the model of BL, consisting of multiple oscillating plasmoids, can explain the visible radiation of such a plasma structure. This radiation is due to the decay of plasma oscillations in some small fraction of oscillating kernels, whereas the majority of them are stable.

The results of the present work can be used to explain the formation of a plasmoid in natural conditions. Let us suppose that during a thunderstorm a high tension, related to a linear lightning strike, is applied to a small object with the appropriate size. It can be a natural spike, a piece of dust or soot etc. Then this object turns out to be embedded in plasma with $n_0 \sim (10^{19} - 10^{20})\ \text{cm}^{-3}$. Moreover, this object will be a source of divergent waves, where the electron density can rise almost one order of magnitude~\citep{Gor10,Bul12} and reach the value required to excite an axial or a spherical Langmuir soliton described in our work. Note that the plasma heating up to  $\sim 10^{6}\ \text{K}$ is also possible in the vicinity of this object. We remind that the typical electron temperature in a linear lightning bolt is~\citep{Rak98} $T_e \approx 3 \times 10^{4}\ \text{K}$. Note that at high electron temperature, atoms can be multiply ionized. It will also increase the electron density.

It should be mentioned that ions are unlikely to be involved in the divergent waves because of their low mobility. Thus the density of ions should keep the initial value, i.e. be lower than that of electrons, in some region near the object. It can explain the formation of a cavern (see Eq.~\eqref{eq:nniondens} and the work by~\citet{Zak72}).

%Taking into account the conservative estimates for the background number density, $n_0 \sim 10^{21}\ \text{cm}^{-3}$, and the electron temperature, $T_{e}=10^{6}\ \text{K}$, necessary for the existence of a plasma structure, as well as its small energy, one cannot directly identify a plasmoid, described in frames of our model, with BL. We remind that the Loschmidt's number is $n_{l}=2.7\times10^{19}\ \text{cm}^{-3}$ and the typical electron temperature in a linear lightning bolt is~\cite{Rak98} $T_e \approx 3 \times 10^{4}\ \text{K}$.

Therefore we may consider a Langmuir soliton, which involves the interaction of induced EDM of ions with the oscillating electric field, as proto-BL, i.e. an atmospheric plasma structure at its initial stages of evolution. Under certain conditions, when other (maybe quantum) nonlinear effects become important, this proto-BL can then be transformed into a glowing object identified as BL.

We also mention that besides the explanation of the formation of natural plasmoids, the model of radial plasma oscillations, involving diatomic ions,
can be used for the interpretation of the results of the experiments performed by~\citet{Kli94,Dim94,KirSavKad95}, where spherical luminous structures were obtained in electric discharges in liquid nitrogen. Indeed, in case of a plasmoid in liquid nitrogen we can take that $T_i = 77\thinspace\text{K}$ and $n_0 \approx 3.44 \times 10^{21}\thinspace\text{cm}^{-3}$, which corresponds to 10\% ionization. To satisfy conditions in Eqs.~\eqref{eq:xidef}-\eqref{eq:nniondens} one should require that $T_e \sim 10^4\thinspace\text{K}$. Such electron temperatures are achievable in laboratory plasmas. It makes rather plausible interpretation of the results of~\citet{Kli94,Dim94,KirSavKad95} in frames of our model.

\section{\label{sec:CONCL}Conclusion}

In the present work we have studied stable spatial plasma structures
possessing axial and spherical symmetry. The stability of soliton-like
plasmoids against the collapse is provided by the defocusing interaction
of the induced EDM of ions with the rapidly oscillating electric
field. Note that an ion was supposed to be diatomic and nonpolar. In Sec.~\ref{sec:NLLO}, starting from the complete set of plasma
hydrodynamic equations we have
derived the basic nonlinear equations for the envelope of the electric
field and for the perturbation of the ion density (see Eqs.~\eqref{eq:ZakheqE} and~\eqref{eq:Zakheqn}).
In Sec.~\ref{sec:CQNLSE} we have reduced these equations to a single
cubic-quintic NLSE~\eqref{eq:NLSEpsi}, written in the dimensionless
variables. Then Eq.~\eqref{eq:NLSEpsi} was solved
numerically.
Applying VKC, we have found that mainly stable spatial solitons can
exist in 2D case, whereas in 3D case both
stable and unstable solitons are present.

It should be noted that VKC was originally formulated by~\citet{KuzRubZak86} for a NLSE for a scalar ``wave function'' $\Psi$. In 2D or 3D cases, the kinetic term of such an equation contains the Laplace operator of $\Psi$, which reads $\Psi'' + \tfrac{d-1}{r}\Psi'$ for the function depending only on the radial coordinate. The kinetic term of NLSE~\eqref{eq:NLSEpsi}, derived in our work, has different structure. Nevertheless, as shown by~\citet{DavYakZal05}, the application of VKC to the analysis of NLSE analogous to Eq.~\eqref{eq:NLSEpsi} is justified.

In Sec.~\ref{sec:APPL} we have discussed a possible application
of the obtained results to the studies of a long-lived
atmospheric plasma structure, BL. It was found that in a realistic
atmospheric plasma, mainly composed of $\mathrm{N}_{2}^{+}$ and $\mathrm{O}_{2}^{+}$
ions, one may expect that the initial stages of BL evolution can be described in frames of the plasmoids model developed in
the present work. We have also suggested one of the possible ways of the creation of
this kind of plasma structures in natural conditions. Besides the description of natural plasmoids, in Sec.~\ref{sec:APPL} we have considered the application of our model for the interpretation of the results of experiments where plasma structures were generated in liquid nitrogen.

As we mentioned in Sec.~\ref{INTR}, besides the mechanism proposed in the present work, there are other ways to arrest the collapse of a Langmuir wave. For example, it was shown by~\citet{Kuz76,SkoHaa80,DavYakZal05} that the electron-electron nonlinear interaction prevents the soliton collapse. It should be noticed that in this case higher harmonics are generated in the system. However, a plasma structure, supported by this nonlinearity, was demonstrated by~\citet{Dvo11} to exist only in the upper ionosphere, where the density is very low. We remind that the majority of the BL observations indicate that this phenomenon happens in the lower troposphere~\citep{BycNikDij10}.

Note that in our work we have used polarizabilities of a neutral nitrogen molecule rather than of $\mathrm{N}_{2}^{+}$, which in fact can differ significantly. Nowadays no measurements of $\mathrm{N}_{2}^{+}$ polarizabilities
have been made. This fact can change the estimate of the critical background density, necessary for the plasmoid existence, towards its reduction.

\appendix

\section{\label{sec:POLARIZ}Polarization of a nonpolar diatomic gas in an
external electric field}

%In this Appendix we shall derive the permittivity of a gas of a nonpolar
%diatomic molecules in an external electric field.

The collective response of plasma results in the concept of ``dressed''
particles. It means that the Coulomb interaction of a charged particle
is replaced by the Debye-H\"{u}ckel potential, $\tfrac{1}{r}\exp\left(-\tfrac{r}{r_{\mathrm{D}}}\right)$.
If the plasma density is quite high, i.e. the Debye length is
short, we may suppose that the electric charge of an ion is quite
perfectly screened by surrounding electrons. In this case the Hamiltonian
of a diatomic ion possessing an axial symmetry in an external electric
field reads
\begin{equation}
  \mathcal{H}=\frac{\mathbf{M}_{\bot}^{2}}{2I}+U,
\end{equation}
where $\mathbf{M}_{\perp}$ is the vector of the angular momentum
of an ion perpendicular to the ion axis, $I$ is the moment of inertia
of an ion, $U=-(\mathbf{p}\mathbf{E})=-\mathbf{E}^{2}\left(\alpha_{\bot}\sin^{2}\theta+\alpha_{\Vert}\cos^{2}\theta\right)$
is the potential energy of a polarized ion in an external electric
field, and $\theta$ is the angle between the ion axis and the electric
field direction.

On the classical level the thermodynamic properties of a gas can
be calculated on the basis of the canonical partition function,
$Z=z^{\mathcal{N}}$, where $\mathcal{N}$ is the total number of
ions. The reduced partition function $z$ which includes the rotational
degrees of freedom of an ion has the form,
\begin{align}
  z_{\mathrm{rot}} = & \frac{1}{(2\pi\hbar)^{2}}
  \int\mathrm{d}^{2}\mathbf{M}_{\bot}\exp
  \left(
    -\frac{\mathbf{M}_{\bot}^{2}}{2IT_{i}}
  \right)
  \int\mathrm{d}\Omega
  \nonumber
  \\
  &
  \times \exp
  \left[
    \frac{\mathbf{E}^{2}}{T_{i}}
    \left(
      \alpha_{\bot}\sin^{2}\theta+\alpha_{\Vert}\cos^{2}\theta
    \right)
  \right],
\end{align}
where $\mathrm{d}\Omega=2\pi\sin\theta\mathrm{d}\theta$ is the solid
angle differential.

Calculating the integral over the angular momentum components, we
can express $z_{\mathrm{rot}}$ in the following form:
\begin{align}
  z_{\mathrm{rot}} = &
  \frac{2I}{\hbar^{2}}T_{i}\exp
  \left(
    \frac{\mathbf{E}^{2}}{T_{i}}\alpha_{\bot}
  \right)
  z',
  \notag
  \\
  z' = & \int_{0}^{1}\mathrm{d}xe^{\xi x^{2}} =
  1+\frac{\xi}{3}+\frac{\xi^{2}}{10}+\dotsb,\label{eq:zrot}
\end{align}
where $\xi=\mathbf{E}^{2}\Delta\alpha/T_{i}$. The polarization of
the gas in an external electric field can be calculated on the basis
of the expression for the free energy $F=-T_{i}\ln Z$, as
\begin{align}
  \mathbf{P} = & -\frac{1}{V}
  \left(
    \frac{\partial F}{\partial\mathrm{\mathbf{E}}}
  \right)_{T_{i}} 
  \notag
  \\
  & =
  %n_{i}T_{i}\left(\frac{\partial\ln z_{\mathrm{rot}}}{\partial\mathbf{E}}\right)_{T_{i}}=
  2\mathbf{E}n_{i}
  \left(
    \langle\alpha\rangle + \frac{4}{45} \frac{(\Delta\alpha\mathbf{E})^{2}}{T_{i}}
  \right),\label{eq:polariz}
\end{align}
where $V$ is the volume of a gas. Here we account for the decomposition
of $z'$ in Eq.~\eqref{eq:zrot}.

Basing on Eq.~\eqref{eq:polariz} we obtain the permittivity of the
ion component of plasma
\begin{equation}\label{eq:perm}
 \varepsilon = 1 + 8\pi n_{i}
 \left(
   \langle\alpha\rangle + \frac{4}{45}
   \frac{(\Delta\alpha\mathbf{E})^{2}}{T_{i}}
 \right),
\end{equation}
which was used in Sec.~\ref{sec:NLLO} to derive the ponderomotive
force acting on ions. Using Eqs.~\eqref{eq:zrot} and~\eqref{eq:polariz}, one can get the quartic contribution to the permittivity in Eq.~\eqref{eq:perm}. It has the form,
\begin{equation}\label{eq:perm4}
 \Delta \varepsilon = \frac{64}{945} \pi n_i \frac{\Delta\alpha^3 \mathbf{E}^4}{T_{i}^2}.
\end{equation}
The obtained correction to $\varepsilon$ is small. However, using the fact that $\Delta \varepsilon$ in Eq.~\eqref{eq:perm4} is positive as well as the formalism developed in Secs.~\ref{sec:NLLO} and~\ref{sec:CQNLSE}, we get that, if one accounts for this correction, it will defocuse Langmuir waves and further stabilize solitons.

Finally we mention that our calculations are based on the assumption
of the constant electric field whereas in Secs.~\ref{sec:NLLO} and~\ref{sec:CQNLSE}
the electric field is supposed to oscillate with the high frequency
$\sim\omega_{e}$. According to~\citet{AlmBurFly75}, the typical frequency associated with the polarizability of a
molecule is $\sim10^{15}\ \text{Hz}$. This value is several orders of
magnitude greater than plasma frequencies in realistic plasmas. Thus
the approximation of the constant electric field is valid.

\section*{Acknowledgements}
%\begin{acknowledgements}
I am thankful to S.~I.~Dvornikov, A.~I.~Nikitin and J.~T.~Mendon\c{c}a for helpful comments, to V.~Berenguela for providing some references, and to FAPESP (Brazil) for a grant.
%\end{acknowledgements}

\end{document}